\documentclass[12pt]{article}

\newcommand{\Z}{\ensuremath{\sf{Z\!\!Z}}}

\begin{document}

\title{Quantum Return Probability for Substitution Potentials}
\author{C\'esar R. de Oliveira\thanks{e-mail: oliveira@dm.ufscar.br} {\small and} Giancarlo Q. Pellegrino\thanks{e-mail:
gian@dm.ufscar.br}\\
\vspace{-0.0cm}
\it \small Departamento de Matem\'{a}tica -- UFSCar, 
\vspace{-0.0cm}
 \it S\~{a}o Carlos, SP, 13560-970 Brazil }

\vspace{1cm}
\date{ \today} 
\maketitle

\vspace{2cm}

\begin{abstract}

We propose an effective exponent ruling the algebraic decay of the average quantum return probability for discrete
Schr\"odinger operators. We compute it for some non-periodic substitution potentials with different degrees of randomness, and do not
find a complete qualitative agreement with the spectral type of the substitution sequences themselves, i.e., more random the sequence
smaller such exponent.

\end{abstract}

PACS numbers: 05.45.Pq, 02.30.-f, 71.30.+h, 71.55.Jv

\clearpage

Anomalous transport in non-periodic structures is due to intricate quantum interferences which may also lead to localization of wave
functions. Another possibility is ballistic motion, mainly related to periodic structures. Here we consider transport properties in
nearest-neighbours tight-binding models in~$\Z$ whose general Hamiltonian $H$ is given by
\begin{equation}
\label{hamiltonian} (H\psi)_n=\psi_{n+1}+\psi_{n-1}+\lambda V_n\psi_n,
\end{equation} with~$\lambda>0$ and potentials $V=(V_n)_{n\in\Z}$ generated by some non-periodic substitution sequences.

Among the characterizations of (de)localization and transport we single out the (average) moments of the ``position'' operator
\begin{equation}
\label{moments} m_\alpha(T)=\frac{1}{T}\int_0^T dt\sum_{n=-\infty}^\infty |n-n_0|^\alpha |\psi_n(t)|^2,\;\;
\alpha\neq 0,
\end{equation} and the (average) return probability
\begin{equation}
\label{return} C(T)=\frac{1}{T}\int_0^T dt\; |\psi_{n_0}(t)|^2 .
\end{equation}  In relations~(\ref{moments})~and~(\ref{return}) it is implicitly assumed that the initial condition is
$\psi_n=\delta_{n,n_0}$. Both quantities
$m_\alpha(T)$ and $C(T)$ have strong physical appeal and in some cases are attainable to theoretical and numerical investigations. We
notice that the return probability was one of the first quantities considered in the seminal paper by Anderson on localization in
disordered structures~\cite{A}. 

It has been found that for large $T$~\cite{G1,GM1,KPG,H,BCM,L} 
\begin{equation}
\label{exponents} m_\alpha (T)\sim T^{\alpha\beta(\alpha)}\;\; {\rm and }\;\; C(T)\sim T^{-\Delta}.
\end{equation}  Localization should be characterized by vanishing exponents $\beta$ and $\Delta$, ballistic motion by
$\beta(\alpha)=\Delta=1$, while anomalous transport by $0<\beta,\Delta<1$. Notice that $\beta(2)$ is related to the direct conductivity
via the anomalous Drude formula~\cite{SBB1,SBB2}. 

In this work we consider potentials $V$ in~(\ref{hamiltonian}) generated by some substitution sequences and compute the exponent for
the decay of the return probability as a function of the degree of randomness of those sequences. As it will be seen later, in
most cases~$\Delta$ can not be obtained  directly from numerical integration of the time-dependent Schr\"odinger equation; namely,
the standard fitting procedure given by equation~(\ref{exponents}) works only for the well-investigated case of
Fibonacci potentials (to be defined below).  We then propose an
alternative approach, based on the energy spectral decomposition of the initial quantum state, which not only is able to retrieve
the known Fibonacci results but permits us to exhibit an effective exponent also for other non-periodic substitution potentials such
as Thue-Morse, Rudin-Shapiro, paperfolding and period doubling (see below for their precise descriptions). The main conclusion will
be that  there is no perfect correspondence of the exponent ruling the algebraic decay of the return probability and the degree of
randomness of the own substitution sequences, as one expects based on general considerations (i.e., more random the sequence smaller
such exponent). Now we proceed to the details of the points just outlined, including a justification for the choice of~$\Delta$ as
our exponent of interest.

The RAGE theorem and Wiener lemma give direct physical meaning to the standard types of spectra, i.e., point and continuous
(absolutely and singular) in the sense that one of their corollaries is that for continuous spectra
$\beta(\alpha)>0$ and $\;\Delta>0$, while for point spectra
$\Delta=0$; it is left room for $\beta>0$ even for point spectra due to the tail of eigenfunctions and domain
intricateness~\cite{dRJLS1,dRJLS2}. The determination of such exponents is a quantitative step from RAGE and Wiener lemma which, by its
turn, is still related to deep spectral quantities, i.e., generalized dimensions of the (positive) spectral measures associated to the
initial state~$\psi$. It has been rigorously proven that $\beta(\alpha)$ is bounded from below by the information dimension
$D_1$~\cite{G1,GM1} (all dimensions are related to the corresponding spectral measure) and also conjectured that
$\beta(2)\approx D_0$ ($D_0$ denotes the fractal dimension of the spectrum). In an interesting paper Guarneri and Mantica~\cite{GM2}
have presented examples of homogeneous fractal spectral measures, i.e., with generalized dimensions
$D_q=D_0$ for any $q$, for which
$\beta(\alpha)$ is not constant and no simple exact relation seems to hold between the thermodynamics of the spectrum and the exponent
$\beta$, so that ``multiscaling does not require multifractality''~\cite{GM2}; this was called quantum intermitency in~\cite{GM2,M1}.
See~\cite{BSB} for some recent results on a particular class of systems and other references.

On the other hand, it was rigorously proven~\cite{H,BCM} that the exponent~$\Delta$ ruling the algebraic decay of the return
probability equals the correlation dimension~$D_2$. We note that such relation supposes the limit defining the dimension~$D_2$
does exist. See also~\cite{KPG} where this relation was first proposed in the context of anomalous diffusion. Therefore we have
selected the return probability, its corresponding exponent~$\Delta$ and correlation dimension~$D_2$, as the main tools for analyzing
our systems.

Relevant examples of anomalous diffusion are generated by almost-periodic potentials $V$; an important class of such potentials is
given by $V$ induced by non-periodic (primitive) substitution sequences~\cite{Q,AG}. These sequences form a convenient laboratory for
the study of anomalous transport since in all rigorously analyzed cases they generate singular continuous spectra for the
tight-binding model~(\ref{hamiltonian}), although the own spectral types of substitution sequences are not equal; for example, 
Fibonacci ({\sc Fcc}), paperfolding ({\sc PF}) and period doubling ({\sc PD}) substitution sequences have point autocorrelation
measures, Thue-Morse ({\sc TM}) has singular continuous autocorrelation measure, and the autocorrelation measure of Rudin-Shapiro
({\sc  RS}) substitution sequence is absolutely continuous. Although all these sequences
are almost periodic, their spectral properties characterize them qualitatively from ``ordered to random'' since periodic and
quasiperiodic sequences have pure point autocorrelation measures (as {\sc Fcc, PF} and {\sc PD} do), whereas independent random
sequences have absolutely continuous autocorrelation measures (as {\sc RS} does). The {\sc TM} sequence lies in an intermediate place.
It is also worth noting that all these sequences give rise to strictly ergodic dynamical systems with zero topological and generalized
entropy~\cite{Q,B}.

Due to  different degrees of randomness of the substitution sequences, it was expected differences in the spectral properties of the
corresponding tight-binding Hamiltonians~(\ref{hamiltonian}), but as already commented above all rigorously studied cases have
presented singular continuous spectrum~\cite{BBG,BG1,BG2,Su,HKS,HM} (the RS case is an important open problem~\cite{BG1,HKS,Al};
the {\sc PF} case is also open). Our main goal in this Letter is to investigate whether the spectral character of the sequence
generating the potential is responsible for different physics through details of the return probability behaviour. To this end we
consider potentials
$V$ in~(\ref{hamiltonian}) generated by the five non-periodic sequences {\sc Fcc, PF, PD, TM} and {\sc RS}. $\Delta$ can be computed 
either from numerical integration of the time-dependent Schr\"odinger equation, and then fitting a straight line on $\log
C(t)\times\log t$, or directly computing
$D_2$ from its definition (which also involves a straight line fitting---see below). However, as already anticipated, we have faced
problems in linear fittings in both procedures (except for the well-investigated case of  {\sc Fcc} potentials) and we propose a
pragmatic approach to get such exponents which is able to recover the {\sc Fcc} known results.

Now we present the rules describing the sequences we use to generate $V$. {\sc Fcc, PD} and {\sc TM} sequences are constructed with an
alphabet of two letters $\{a,b\}$ through the substitutions 
\[
 a\rightarrow ab, \;b\rightarrow a\; {\rm(Fcc), \;\;\;}\;\;\; a\rightarrow ab,
\;b\rightarrow ba \;\;\;{\rm(TM),}
\]
\[
 a\rightarrow ab, \;b\rightarrow aa\; {\rm(PD).}
\] Beginning with $a$ and applying successively the substitution rules, non-periodic sequences are obtained; e.g., the Thue-Morse
sequence is given by
\[ abbabaabbaababba\cdots
\]
 The {\sc RS} and {\sc PF} sequences can be obtained with an alphabet of four letters $\{a',b',c',d'\}$, the substitutions 
\[ a'\rightarrow a'b', \;\;\;b'\rightarrow a'c',\;\;\; c'\to d'b',\;\;\; d'\to d'c' \;\;\;\;{\rm(RS)},
\]
\[ a'\rightarrow a'b', \;\;\;b'\rightarrow c'b',\;\;\; c'\to a'd',\;\;\; d'\to c'd'  \;\;\;\;{\rm(PF)},
\] and then the identifications $a',b'\to a$ and $c',d'\to b$ in both cases; the first elements of the RS sequence are
\[ aaabaabaaaabbb\cdots
\]

We then use these substitution sequences to define our potentials $V$; we take 
$V_n=0$ if the $n$-th letter of the sequence is $a$ and $V_n=1$ in case it is $b$. There are standard ways to extend the potential for
negative values of~$n$~\cite{BG2,HKS}, but we avoid such issue by taking a finite sample of $N$ sites, with $n\geq0$, and using the
initial wavefunction~$\psi_n=\delta_{N/2,n_0}$. In this way we construct the almost-periodic substitution potentials $\lambda V$ and
investigate~$\Delta$ as function of the degree of randomness of the potential and its intensity
$\lambda$.

It is known that {\sc Fcc, PD} and {\sc TM} generate potentials whose spectra of~(\ref{hamiltonian}) are singular continuous for all
$\lambda\neq0$. The case of {\sc RS} has been numerically investigated in~\cite{DJR1,DJR2} indicating point spectrum for
$\lambda> 2$ and mixed spectrum, i.e., point and singular continuous, for $0<\lambda\leq2$ (notice we use a  scale for the potential
values which is different from~\cite{DJR1}). For Hamiltonian~(\ref{hamiltonian}) with {\sc PF} potential it is only known that its
spectrum has no absolutely continuous component, since it is primitive~\cite{HKS}; from a rigorous point of view the lack/presence of
eigenvalues in this case is also an open question.

The case of {\sc Fcc} Hamiltonian has also been considered in~\cite{KPG} and a good agreement between the value of $\Delta$ from
numerical integration of Schr\"odinger equation and $D_2$ was found. Let's recall the definition of $D_2$ associated to a spectral
measure~$\mu$ and how it is usually estimated~\cite{KPG,BCM}. For $\varepsilon>0$ let $B_\varepsilon (x)$ denote the open ball of
centre $x$ and radius~$\varepsilon/2$ and set
$$
\gamma(\varepsilon)=\int\mu(B_\varepsilon(x))d\mu(x);
$$  the correlation dimension of~$\mu$ is given by the limit
\begin{equation}
\label{D2} D_2=\lim_{\varepsilon\to 0} \frac{\log\gamma(\varepsilon)}{\log\varepsilon}.
\end{equation} If this limit does not exist one defines~$D_2^+$ and~$D_2^-$ via $\limsup$ and $\liminf$, respectively. The latter
remark is important here since we have found numerical indications that the limit in the definition~(\ref{D2})  does not exist  for
general substitution potentials. In numerical practice we have a finite basis approximation for~(\ref{hamiltonian}) whose spectrum is
composed of eigenvalues~$\chi_k$; then we divide the energy range into boxes~$B_j$ of length~$\varepsilon$, approximate
\begin{equation}
\label{gamaTil}
\gamma(\varepsilon)\approx \gamma^*(\varepsilon) = \sum_j (\sum_{\chi_k\in B_j} |a_k|^2)^2
\end{equation} and get $D_2$ from the linear fitting of $\log \gamma^*(\varepsilon)\times\log\varepsilon$. $a_k$ is the projection of
the initial wavefunction~$\psi$ on the eigenvector with eigenvalue~$\chi_k$. We have used these procedures to recover~$D_2$
and~$\Delta$ found in~\cite{KPG} for the~{\sc Fcc} case as illustrated in figure~1. However such techniques do not work for
substitution potentials distinct from {\sc Fcc} (as far as we have checked), since no clear region with linear behavior is found in
the plots
$\log \gamma^*(\varepsilon)\times\log\varepsilon$ and $\log C(t)\times\log t$, as exemplified in figure~2 for the {\sc PD} potential
with $\lambda=1.6$. We suspect this behaviour is an indication that the limits defining the scale exponents $D_2$ and $\Delta$ are not
well defined in such situations; then we propose a pragmatic approach to extract effective exponents $D_2$ by
selecting a particular value~$\varepsilon^*$ of $\varepsilon$. Before presenting our approach we stress we have also tried to get
well-defined exponents by site averaging on samples beginning at locations
$0,1\times10^4,2\times10^4,\cdots,5\times10^4$, but quite similar behaviours were found.

We begin our argument with the remark that if $\varepsilon$ is smaller than the least eigenvalue spacing (we just ignore the
possibility of degenerate eigenvalues in this argument) then 
$$
\gamma^*(\varepsilon)= \sum_k  |a_k|^4
$$ which resembles the so-called inverse participation ratio (which does not depend on~$\varepsilon$); this also gives a
physical interpretation  for
$D_2$. As a naive first guess for an effective exponent one could try to use $\sum_k  |a_k|^4$ instead of $\gamma^*(\varepsilon)$,
but the exact value of
$\varepsilon$ to be used in an approximation to~(\ref{D2}) is not clear at all. The theoretical determination of
$D_2$ involves the limit
$\varepsilon\to 0$; finite basis approximations preclude this limit and also too small values of $\varepsilon$ are meaningless,
despite the inverse participation ratio interpretation. For sufficient small values of
$\varepsilon$ we have
\begin{equation}
\label{divisao}
\frac{\gamma^*(\varepsilon)}{ \sum_k  |a_k|^4}\approx 1;
\end{equation}  we suggest to pick $\varepsilon^*$ as the smallest value of $\varepsilon$ such that the l.h.s.  of~(\ref{divisao})
considerably deviates from~$1$, so still keeping track of $\gamma^*(\varepsilon)$ and also the inverse participation ratio
interpretation in operation. Then we estimate the effective~$D_2$ as~$D^*_2$ given by
\begin{equation}
\label{epsStar}D^*_2= \frac{\log\gamma^*(\varepsilon^*)}{\log\varepsilon^*}.
\end{equation} 
Let's be more precise on how we have picked up~$\varepsilon^*$ in practice. By using double precision (16 digits) in our code, we
adopted that after diagonalization we can numerically resolve the spectral quantities with 8 digits, i.e., half of the number of
digits of the code precision, so that~$\varepsilon^*$  is given by the smallest value of~$\varepsilon$ such that
$\gamma^*(\varepsilon)-\sum_k  |a_k|^4\geq10^{-8}$ or, equivalently, the smallest~$\varepsilon$ such that
$$
|\frac{\gamma^*(\varepsilon)}{ \sum_k  |a_k|^4}-1|\geq\frac{10^{-8}}{\sum_k  |a_k|^4}.
$$ 
We remark that in most cases~$\varepsilon^*$ can also be obtained directly from visual inspection, as in figure~2b, and the precise
value~$10^{-8}$ is not so relevant since in general~$\gamma^*(\varepsilon)$ has a pronounced jump at~$\varepsilon=\varepsilon^*$.

We have tested our approach in the {\sc Fcc} case  and have got very good agreement with the computed values
of
$D_2$ from our linear fittings and the values reported in~\cite{KPG}. In figure~2b we show a typical curve used to estimate
$\varepsilon^*$, and in figure~3 we compare the values of the exponents $\Delta$ and $D_2$ as calculated in figure~1 and also the
matching values of $D^*_2$ from equation~(\ref{epsStar}) for the {\sc Fcc} case. From now on we use this procedure to estimate the
exponents~$D^*_2$ for the other substitution Hamiltonians~(\ref{hamiltonian}) considered here.

Its now time to discuss our numerical results and details of their  implementations. The return probability $C(T)$ was calculated by
direct diagonalization of the Hamiltonian and eigenfunction expansion of the initial state; we used bases of size~$N\approx
1\times10^3$ and checked some results with bases of size~$N\approx 2\times10^3$. The initial condition was always concentrated on the
centre of the basis~$n_0$ and we have followed its time evolution until time~$T_f$ for which the modulus of the amplitude at one of the
border sites reaches~$1\times10^{-6}$. For the calculation of~$D_2$ we have considered subdivisions of the spectrum in subintervals of
size~$\varepsilon$ ranging from the least eigenvalues spacing (we disregarded multiple eigenvalues) up to~$10^{-2}$. We could seldom
conceive a linear behaviour in such~$\log-\log$ plots in both~$D_2$ and~$\Delta$ cases in order to extract faithful exponents, so that
we were left with the task of finding~$\varepsilon^*$ and computing only~$D_2^*$.

In figure~4 we present a summary of our main numerical results, i.e., the values of~$D_2^*$ for some substitution potentials as
function of the potential intensity~$\lambda$. Since we are not aware of any complete rigorous spectral classification for
Hamiltonian~(\ref{hamiltonian}) in the cases of~{\sc RS} and {\sc PF} sequences, we have also used~$T_f$ as indication of any
possible (de)localization transition; this is the reason for the restriction of the {\sc RS} case to~$\lambda\leq1.7$; for all
substitution sequences we have found~$T_f\leq10^3$ for~$\lambda\leq2$, but for {\sc RS}~$T_f$ jumps from~$T_f\approx10^3$
for~$\lambda=1.7$ to ~$T_f\approx10^5$  for~$\lambda=1.8$, which characterizes absence (at least numerically) of extended states. If
non-localized states are present their ``amounts'' suffer a drastic reduction at~$\lambda\approx1.8$ so that we have not detected
them. Recall that in~\cite{DJR1,DJR2} it is argued that all states of {\sc RS} Hamiltonian should be localized for~$\lambda>2.0$. 

Since no such sharp transition in~$T_f$ was found for the {\sc PF} Hamiltonian, its values of~$D_2^*$ are close to the corresponding
values for {\sc PD}, and both substitution sequences have point autocorrelation measures, we conjecture that the {\sc PF}
Hamiltonian~(\ref{hamiltonian}) has singular continuous spectrum for~$0\neq|\lambda|\leq 2$ (maybe also for any $\lambda\neq 0$) as
{\sc PD} Hamiltonian does~\cite{BBG}.

Besides the above conjecture we see from figure~4 that for all sequences the exponent~$D_2^*$ decreases as~$\lambda$ increases (as
physically expected). Since different exponents were found despite the proven singular continuous spectra of {\sc Fcc, TM} and {\sc
PD} Hamiltonians, we see that~$D_2^*$ is able to discern these operators. The
values of~$D_2^*$ for the {\sc RS} are very close to the {\sc TM} case, but in principle one would not expect this since the
autocorrelation measure of the {\sc RS} sequence is Lebesgue measure (the same for random sequences). Notice that the spectral
classification of the underlying sequence generating the potential does not reflect exactly in~$D_2^*$, since these values for {\sc
PF} and {\sc PD} are below the corresponding ones for {\sc TM} and {\sc RS}, the latter sequences being considered ``more random''
than the former ones. Notice, however, that for the most popular sequences {\sc Fcc, TM} and {\sc RS} we have found agreements with the
classification through the sequence spectral type, i.e., this order implying decreasing values of~$D_2^*$. 

The higher values of the exponents for {\sc RS} compared to {\sc PD} indicates that the presence of extended states mixed with
numerically found localized states~\cite{DJR1,DJR2} does not necessarily imply lower exponents~$D_2^*$.

Summing up, we have found that for substitution potentials in general the dynamical exponent~$\Delta$ and the correlation
dimension~$D_2$ are difficult to be obtained from direct linear fittings (at least with the basis sizes we used; we suspect this is a
consequence of the quantum intermitency and multiscaling in time of the dynamics~\cite{GM2,M1}) and we have
proposed~$D_2^*$ as an effective exponent, which has recovered~$D_2$ in the cases it can be directly obtained. We then
computed~$D_2^*$ for some substitution potentials and have not found a complete qualitative agreement with the spectral type of the
substitution sequences themselves, i.e., more random the sequence smaller~$D_2$. Only for {\sc RS} we have got indications of a
spectral transition from extended (critical) to localized states, although  its values of~$D_2^*$ are higher than those for {\sc PD}
and {\sc PF} cases.

\subsubsection*{Acknowledgments} {\small CRO was partially supported by CNPq (Brazil); discussions with U. Grimm  at
the Max Planck Institute for the Physics of Complex Systems (Dresden, Germany) are acknowledged. GQP thanks the support by FAPESP
(Brazil).}

\vspace{1.0 cm}

\clearpage

\section*{Figure Captions}
\vspace{1cm} {\large{\bf Figure 1}}

a) Log-log (base 10) of the return probability (dashed line) $C(t)$ for the {\sc Fcc} potential with $\lambda=1.0$ versus time. The
slope of the straight line fitting (full line) corresponding to $\Delta$ is indicated.

(b) Log-log (base 10) of $\gamma^*$ (dashed line) for the {\sc Fcc} potential with $\lambda=1.0$ versus $\varepsilon$. The slope of the
straight line fitting (full line) corresponding to $D_2$ is indicated.

\vspace{1.4cm}

\noindent{{\large\bf Figure 2}}

Same as in figure 1 but for the {\sc PD} potential with $\lambda=1.6$. The arrow in~b) indicates $\varepsilon^*$; the first point
at left in b) corresponds to the least eigenvalue spacing, for which $\gamma*=\sum_k |a_k|^4$. No
linear fitting is shown.

\vspace{1.4cm}

\noindent{{\large\bf Figure 3}}

 Scaling exponents $D_2$, $D_2^*$ and $\Delta$ for various potential intensities $\lambda$ of the {\sc Fcc} potential.

\vspace{1.4cm}

\noindent{{\large\bf Figure 4}}

 Effective scaling exponents $D_2^*$ as function of $\lambda$  for some substitution potentials.

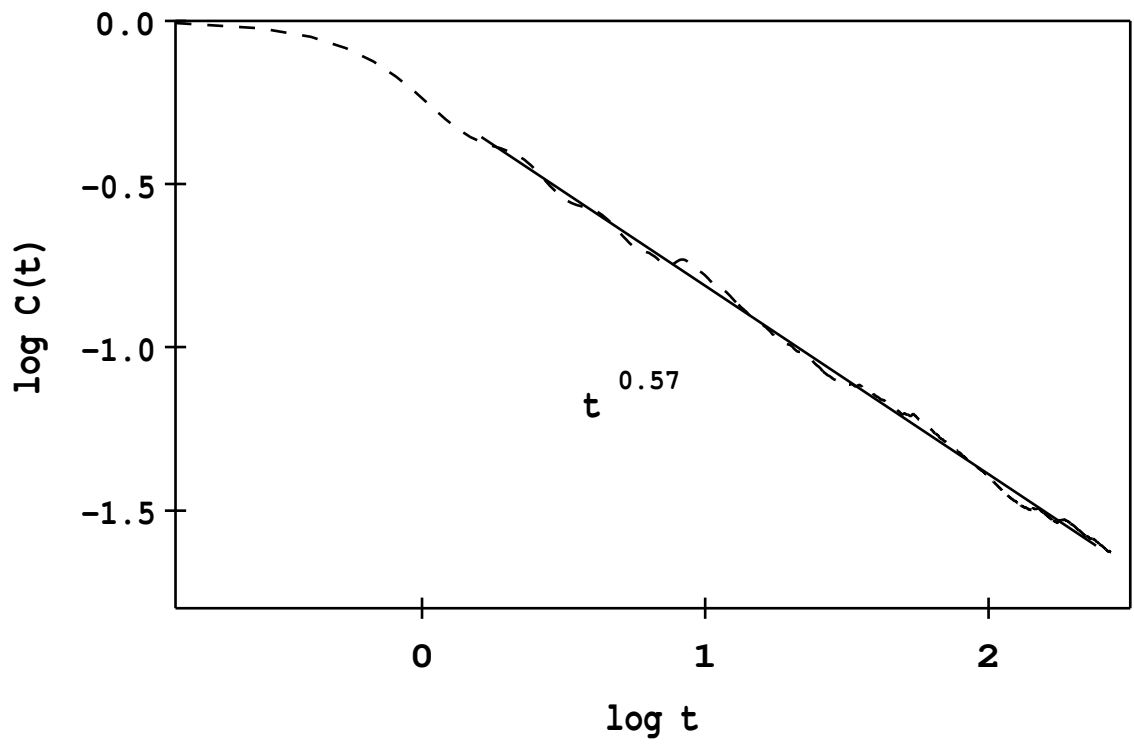
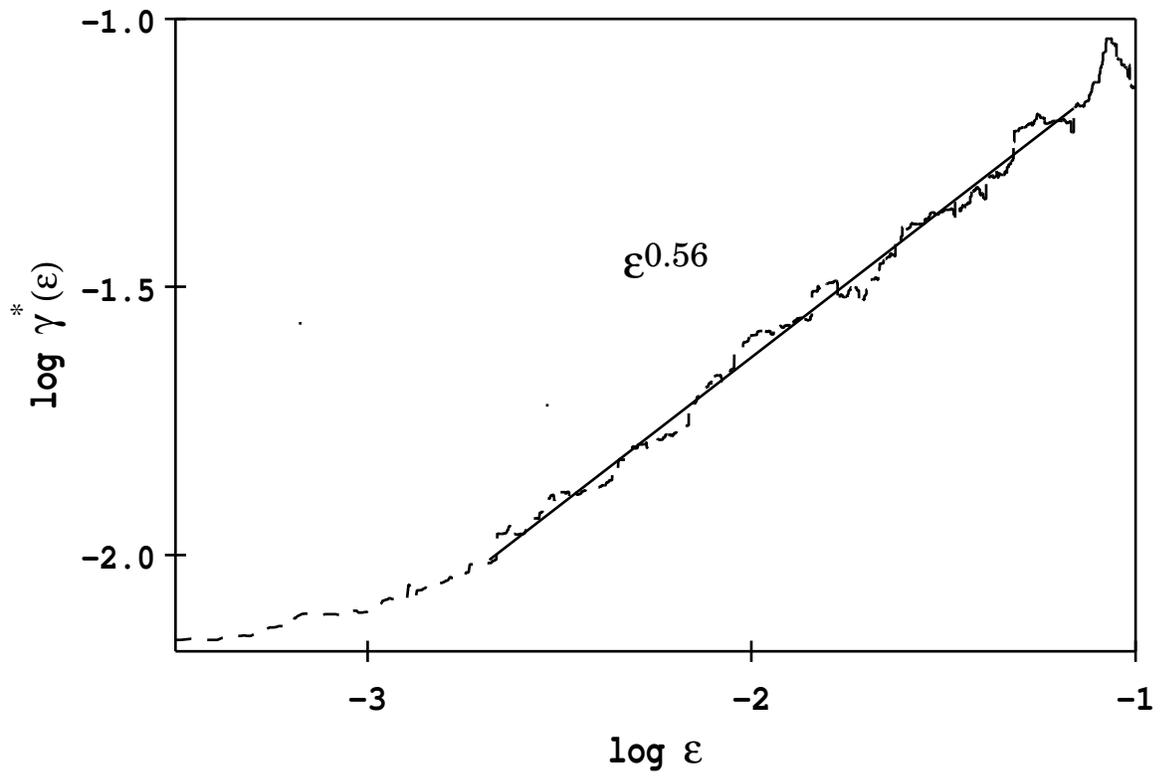
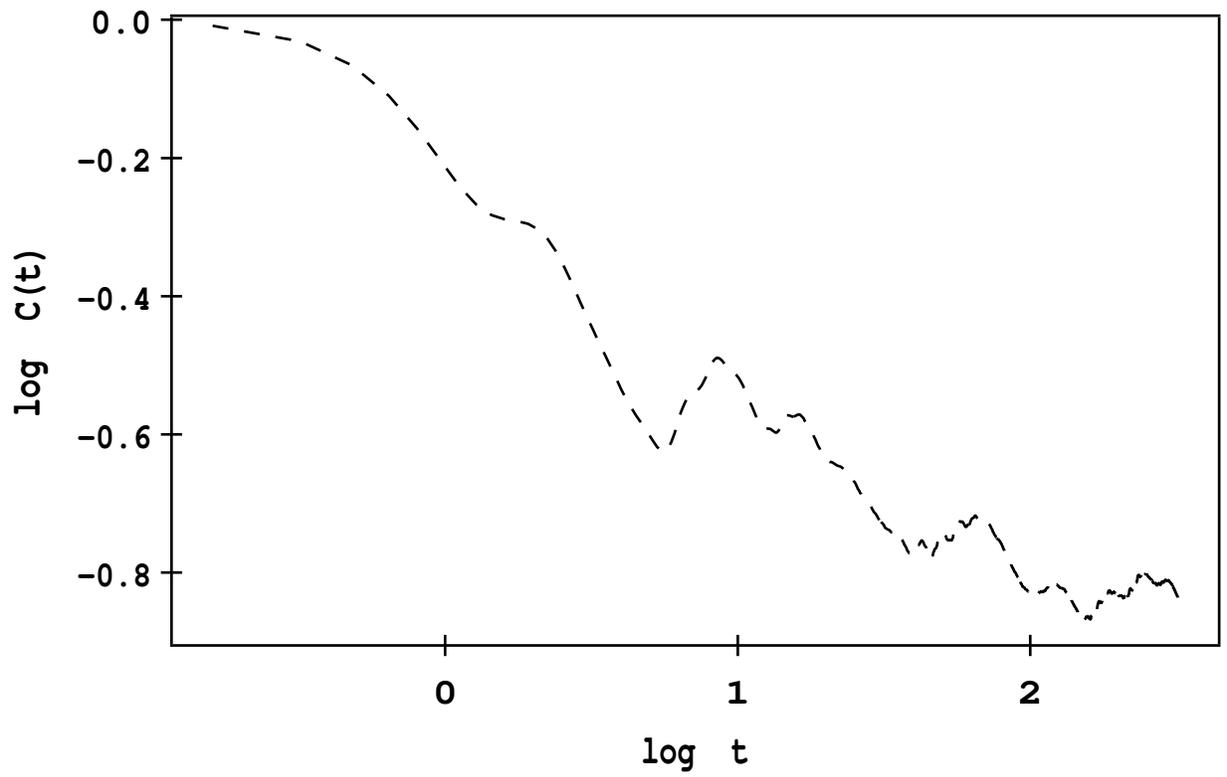
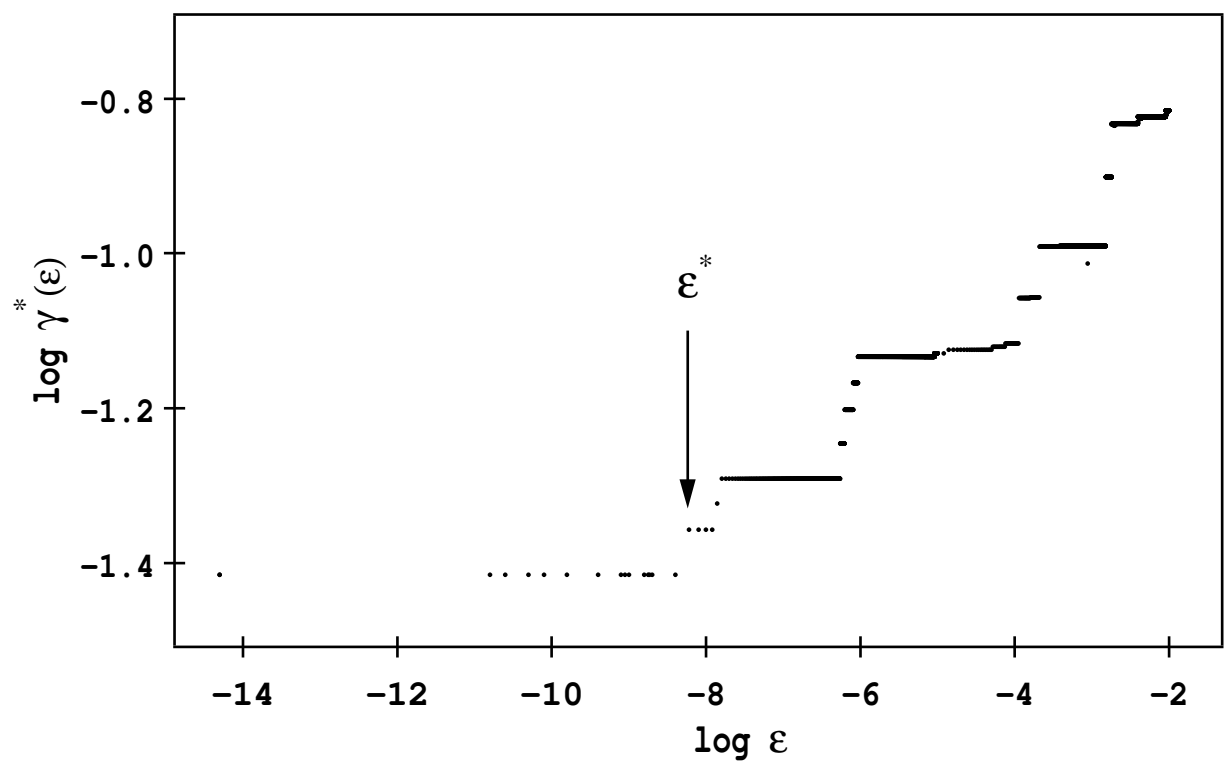
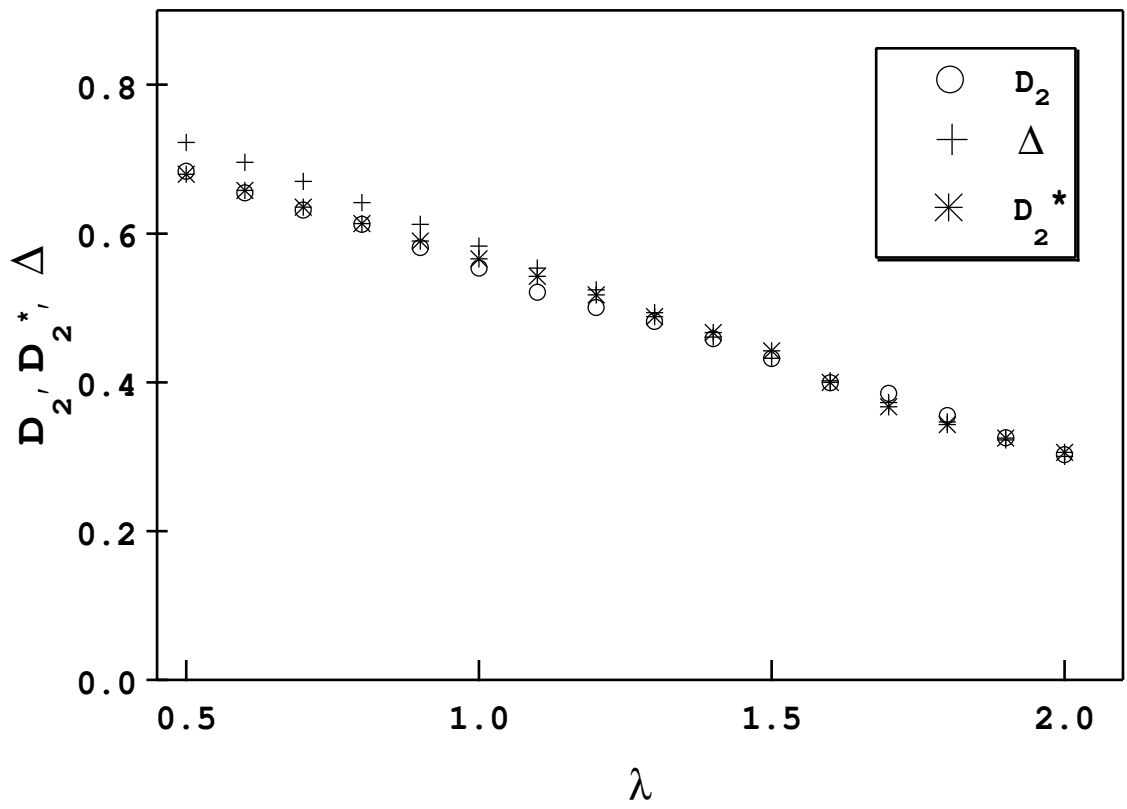
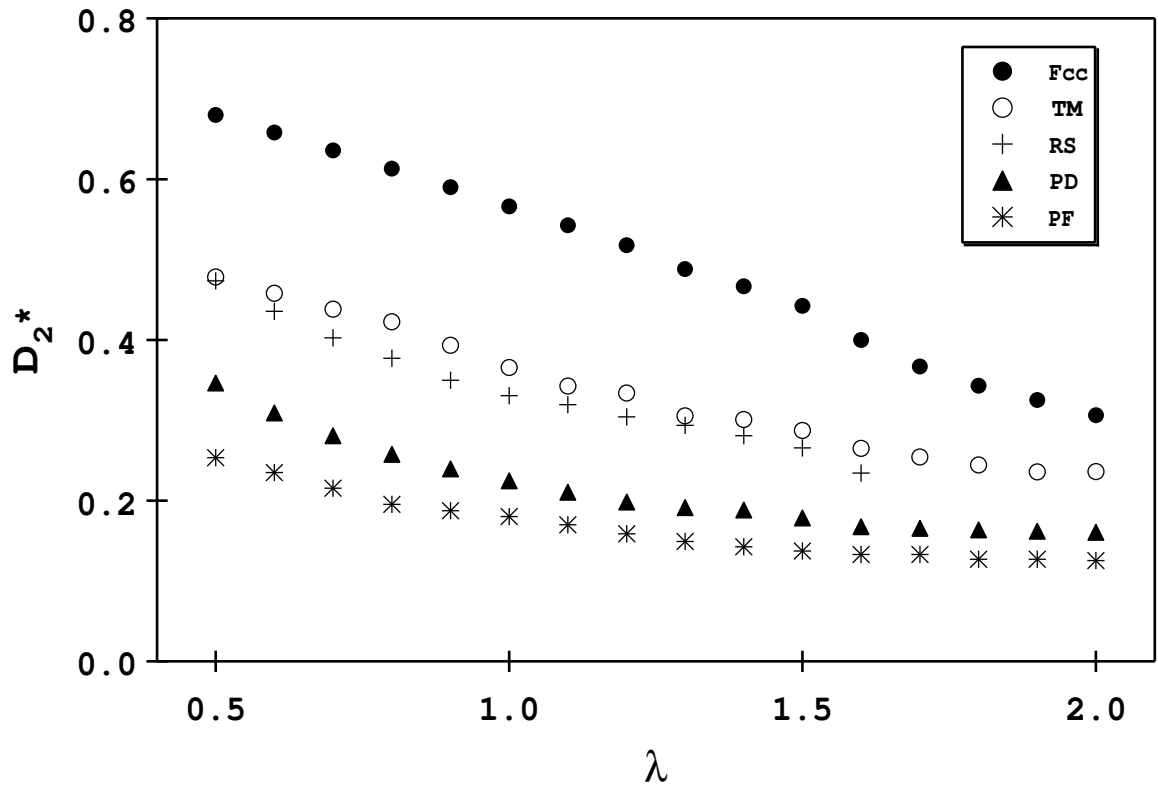
\end{document}